%% file: mbh21.tex
\pdfoutput=1
\documentclass[11pt,preprint]{aastex}

\usepackage{natbib}
\usepackage{amssymb,amsbsy,amsmath}
\usepackage{graphicx}
\usepackage{ifthen,url}
\usepackage{verbatim}
\usepackage{booktabs}
\usepackage{multirow}
\usepackage{colortbl}

\definecolor{mygrey}{rgb}{0.9,0.9,0.9}
\begin{document}

\shorttitle{Constraining the LISA Black Hole Mass Spectrum}
\shortauthors{J. E. Plowman et al.}
\title{Constraining the Black Hole Mass Spectrum with Gravitational Wave Observations I: The Error Kernel}
\author{Joseph E. Plowman\altaffilmark{1}, Daniel C. Jacobs\altaffilmark{1}, Ronald W. Hellings\altaffilmark{1},\\ Shane L. Larson\altaffilmark{2}, Sachiko Tsuruta\altaffilmark{1}}
\altaffiltext{1}{Department of Physics, Montana State University, Bozeman, MT 59717}
\altaffiltext{2}{Department of Physics, Utah State University, Logan, UT 84322}

\begin{abstract}
Many scenarios have been proposed for the origin
of the supermassive black holes (SMBHs) that are found in the
centres of most galaxies. Many of these formation scenarios predict a high-redshift
population of intermediate-mass black holes (IMBHs), with masses $M_\bullet$ in
the range $10^{2} M_{\odot} \lesssim M_{\bullet} \lesssim 10^{5}
M_{\odot}$. A powerful way to observe these IMBHs is via
gravitational waves the black holes emit as they merge. The
statistics of the observed black hole population should, in principle, allow us to 
discriminate between competing astrophysical scenarios for the origin 
and formation of SMBHs. However, 
gravitational wave detectors such as LISA will not be able to
detect all such mergers nor assign precise black hole
parameters to the merger, due to weak gravitational wave signal
strengths. In order to use LISA observations to infer the
statistics of the underlying population, these
errors must be taken into account. We describe here a method for
folding the LISA gravitational wave parameter error estimates into
an `error kernel' designed for use at the population model level.
The effects of this error function are demonstrated by applying it
to several recent models of black hole mergers, and some tentative conclusions 
are made about LISA's ability to test scenarios of
the origin and formation of supermassive black holes.
\end{abstract}


\section{Introduction}\label{sec:Intro}
There is now substantial evidence
\citep[e.g.,][]{KormendyRichstone95, richstone1998sbh, Bend05,
Rees2002, Rees2003} for the existence of supermassive black holes
(SMBHs) in the nuclei of most galaxies, the black hole in our own
galaxy being the best studied and most clearly justified of these
objects. However, the origin of these black holes remains an
unsettled question. In one scenario, the more massive black holes
formed from the merger and coalescence of smaller `seed' black
holes that were created in the very early Universe
\citep[e.g.,][]{Madau_Rees_2001}. Several models
utilizing this process have been proposed and numerically
simulated \citep[e.g.,][]{KauffmannHaehnelt00,VHM03}. In \citep[e.g.,][]{VHM03,tanakahaiman09}, typical seed black holes are the remnants of Population III stars with masses $m_\bullet \sim 100 - 300M_\odot$ formed at high redshift (e.g., $z\sim$ 20 -- 50). Thus, these models
predict an evolving population of intermediate mass black holes
(IMBHs), with masses between $\sim$ 100 to $10^5M_\odot.
$\footnote{Since our studies include both IMBHs, with mass $\sim
100M_\odot$ to $10^5M_\odot$, and SMBHs, with mass $\sim
10^5M_\odot$ to $10^9M_\odot$, we adopt the terminology `massive
black hole' (MBH) to cover the entire range from $\sim 100M_\odot$
to $10^9M_\odot$.} In another version \citep{BVR}, the seed
black holes are formed due to direct collapse of the cores of
pregalactic halos through a `quasi-star' stage, resulting in a more
massive seed population ($ \sim 10^4 - 10^6M_\odot$). This scenario predicts
far fewer IMBHs in the early universe.

Black holes in the IMBH mass range are extremely difficult to
detect with the usual electromagnetic observation techniques,
making it very difficult to verify a particular formation and evolution scenario and, especially, to discriminate between various
models. However, the mergers themselves produce gravitational
waves in the Low Frequency (LF) band, from $10^{-6}$ Hz to
$10^{-1}$ Hz, probed by the proposed LISA mission
\citep{Jennrich04}. Observations of the amplitude, frequency
chirp, and harmonic structure of the gravitational wave waveform
enable both the luminosity distance and the individual masses of the black
holes in the binary system to be determined. LF gravitational wave
observations thus provide a probe of the cosmological spectrum of
black holes, and allow tests of the population models to be made.
In this paper, we investigate how gravitational
wave observations of coalescing massive black hole binaries may be
used to discriminate between models of massive black hole
populations and determine the merger history that has led to the
observed population of SMBHs.

We thus take a slightly different direction, compared to other
recent works investigating LISA detections of black hole
coalescences. Most of these take the somewhat
speculative black hole population models and calculate the number of
coalescences that each model predicts would be observed by LISA.
The scientific goal of these papers has clearly been the observation of the
coalescence itself. In this paper, we turn the problem around and
ask, `What might LISA observations of binary MBH coalescences
tell us about the otherwise uncertain population
models?' 

The organization of the paper is as follows. In Section 2, we discuss 
some of the population models and the motivation behind them.
Section 3 describes the parameters relevant to detection of a black
hole binary and draws a distinction between `population' parameters, 
which are relevant to the population models, and astrophysically uninteresting 
`sample' parameters, which vary randomly for each sample drawn from a 
population. We also review the response of the LISA gravitational wave detector
to massive black hole binary coalescences, with emphasis on the ability of
the detector to determine the parameters of the binary from the gravitational wave
waveform. The bulk of our work is presented in Section 4,
which describes a Monte Carlo calculation of an
`error kernel', which is marginalised over the sample parameters.
This error kernel, $K(\hat\lambda_i,\lambda_i)$, is
the average conditional probability that a source will be detected with population
parameters $\hat\lambda_i$ in the LISA data, given the existence of an 
astrophysical source with population parameters $\lambda_i$. 
The marginalisation over sample parameters sets this apart from previous work,
and the resulting error kernels can be applied directly to model coalescence
rates, producing a new set of coalescence rates as functions of the best-fitting,
`detected', parameter values. Finally, in Section 5, we take several population models from the 
literature and discuss how the error kernel may be used to 
produce a measure of how well the models may be discriminated from 
one another. Although the models we use are incomplete, being based
only on summaries available in the literature, we 
nevertheless draw a few tentative conclusions about LISA's 
ultimate ability to distinguish between black hole population 
models in Section 6.

\section{Astrophysical Populations of Black Hole Binary Systems}\label{sec:astrophysics}

Observations of the high redshift quasar population
\citep{Fan2001,Stern2000,Zheng2000,Becker2001} suggest that a
population of SMBHs has existed since early epochs ($z \sim 6$). 
The local census of SMBHs has been
increasing in recent years \citep{Tremaine2002}, driven by a
growing body of observational evidence linking the mass of SMBHs with
observational properties of their host galaxies.  Early studies
revealed a rough correlation between SMBH mass and the bulge
luminosity of the host galaxy
\citep{KormendyRichstone95,Magorrian98}. A much stronger
correlation was later discovered between the SMBH mass and the
stellar velocity dispersion in the galactic core, the so-called
`$M$-$\sigma$' relation
\citep{Gebhardt2000,FerrareseMerritt2000,Tremaine2002}. The
current best fit to the $M$-$\sigma$ relation
\citep{MerrittFerrareseApJ2001,Tremaine2002} gives the mass of the
central black hole $M_{\bullet}$ as

\begin{equation}
    \log \left(\frac{M_{\bullet}}{M_{\odot}}\right) = 8.13 + 4.02
    \log \left(\frac{\sigma}{200 {\rm km/s}}\right)\ .
    \label{Msigma}
\end{equation}

\noindent The observational data supporting the $M-\sigma$
relation currently spans a mass range from $\sim 10^{5} {\rm
M}_{\odot}$ to $\sim 10^{9} {\rm M}_{\odot}$.

Given this observational evidence for the existence of a SMBH
population, the question arises:  how did these objects come to
be? Several scenarios are proposed \citep[see][]{djorgovski08}:
\begin{enumerate}
\item direct gravitational core collapse of pregalactic dark halos,\label{collapsescenario}
\item growth from seed black holes through merging and coalescences over time,\label{seedscenario}
\item gravitational runaway collapse of dense star clusters,\label{clusterscenario}
\item primordial BH remnants from the big bang.\label{primordialscenario}
\end{enumerate}

In case \ref{collapsescenario} SMBHs can form very early in the Universe through
direct collapse of dark matter halo with mass of $\sim 10^{6} {\rm
M}_{\odot}$ or larger \citep{BrommLoeb02}. In case \ref{seedscenario}, on the other hand,
\citep{Madau_Rees_2001,KauffmannHaehnelt00,VHM03,
tanakahaiman09,BVR}, seed black
holes produced in the early universe are significantly smaller and grow by accretion,
coalescences and merging, leading to the population of SMBHs seen
in the Universe today. In most cases \citep[e.g.,][]{VHM03}
the seed black holes have mass less than $\sim 300M_\odot$, and are 
the remnants of ordinary Population III stars. Detailed stellar 
evolution calculations have recently found, however, that these seed black
holes can be remnants of very massive Population
III stars, referred to as CVMSs
\citep{Ohkuboetal06,Ohkuboetal08,Tsurutaetal07,ohkuboetal09, umedaetaljcap09}. These metal-free CVMSs evolve quickly
and then collapse in the early universe, yielding an IMBH
population with masses in the range of $\sim 500 M_{\odot} -
10,000 M_{\odot}$. In another version \citep{BVR}, even more massive black holes can form from direct core collapse
of mini halos through a quasi-star phase, instead of through
ordinary stellar evolution, leading to a population with masses in the range
 $\sim 10^{4} {\rm M}_{\odot}$ to $\sim 10^{6} {\rm M}_{\odot}$.

The scenario involving direct collapse is not difficult to distinguish
from that involving Population III stars,
because the former predicts only a small number of the more massive 
IMBHs while the latter predicts considerably more IMBHs with a wider 
range of masses and redshifts. In case \ref{clusterscenario}
\citep[e.g.,][]{Ebisuzaketal01, Portegiesetal04a,
Portegiesetal04b}, IMBHs can be formed at any time in dense star
clusters and grow by merging in the given environment. Such a
process can produce a low level population of mergers at all
redshifts. Case \ref{primordialscenario} is theoretically possible, but currently
it will be hard to test by observations.

In this paper, we concentrate on versions of
case \ref{seedscenario}, although the techniques used are general and should 
also apply to discrimination between other 
MBH population scenarios. Specifically, we use the results published in
\cite{Sesana07}, which give coalescence rates for models discussed in \citet{VHM03}, \citet{KBD}, and \citet{BVR}. In these models, the evolution of the seed population through time is carried
out numerically \citep{somervillekollatt, VHM03}. In \cite{VHM03}, for instance, the evolution of a SMBH's host galaxy halo is first computed, starting at the present day and working backwards in
time to $z \sim 20$. Then, the progenitor haloes are seeded with BHs of $150 M_{\odot}$, and the BH merger history is traced forward in time through the halo merger history. Major differences between different versions of this scenario
come from the varying astrophysical assumptions employed - e.g., the
initial conditions, environment, age (the redshifts when the
seeds were produced), seed mass, dynamics, hierarchial growth,
etc., built into the process.

\section{Gravitational wave binaries}\label{sec:binaries}

\subsection{Binary Parameters}\label{sub:parameters}

Each black hole binary may be characterized by a number of
parameters, which are usually  divided into two categories. The
first category is the {\em intrinsic} parameters which have to do
with the local properties of the binary in its rest frame. They
are $m_{1}$, the mass of the primary, $m_2$, the mass of the
secondary, and either the initial orbital separation, $a$,
frequency, $f$ (related to $a$ by Kepler's third law) or time to
coalescence, $t_c$, (related to $a$ by the quadrupole formula)\footnote{In general, the list of intrinsic parameters also
includes the black hole spins and orbital eccentricity. Here,
however, attention is restricted to the simplified case of
non-spinning black holes in circular orbits.}. A set of mass
parameters equivalent to $m_1$ and $m_2$ but more directly related
to the gravitational wave waveform, is the chirp mass ${\cal
M}_{c}=(m_1 m_2)^{3/5}(m_1+m_2)^{-1/5}$ and the symmetric reduced
mass ratio $\eta=(m_1m_2)/(m_1+m_2)^2$. Note that, in the gravitational wave signal, the masses are scaled by the redshift, so that the natural mass variables for gravitational wave data analysis are redshifted. For instance, the redshifted chirp mass is ${\cal M}_c\times (1+z)$, although the reduced mass ratio remains unchanged since it is dimensionless. The remaining parameters fall
into the second category, and are called {\em extrinsic}
parameters. These have to do with the binary's location and
orientation with respect to the LISA constellation. They are the
luminosity distance $D_{L}$ (or equivalently, the redshift $z$),
the principle gravitational wave polarization angle $\psi$, the
binary inclination $\iota$, the sky location angles $\theta$ and
$\phi$, and the initial phase of the binary orbit $\Phi_0$.

Redshift and luminosity distance are used interchangeably as source parameters, with the relationship between them determined from the standard WMAP cosmology ($\Omega_M=0.27$, $\Omega_\mathrm{vac}=0.73$, $\Omega_\mathrm{rad}=0.0$, and a Hubble constant of 71 km/s/Mpc).

Since the predictions of the population studies are given
as functions of masses and redshift, the binary parameters are
best divided into two sets in a {\em different} way, for purposes
of this paper. The first set, consisting of ${\cal M}_{c}$,
$\eta$, and $z$, are what we will call \emph{population}
parameters, since these are the parameters that characterise the
population model predictions. The remaining parameters, $\psi$,
$\iota$, $\theta$, $\phi$, $t_c$, and $\Phi_0$, represent
particular samples drawn from the population model and will be
referred to as \emph{sample} parameters. Sample parameters have
distributions that are essentially stochastic and contain no useful
information about the astrophysical processes which give rise to
the black hole population.

Despite the fact that the sample parameters are not part of the
intrinsic astrophysical model, they can have a dramatic effect on
LISA's source characterisation capabilities because all parameters
must be fit to the data in the process of extracting the
population parameters of interest. A key component of this analysis is therefore to calculate
average LISA error distributions for the population parameters by
averaging over a Monte Carlo ensemble of many sources, each having
randomly-chosen values for the sample parameters. Such error
distributions are referred to as `error kernels'.

\subsection{Gravitational Waves from Binary Systems}\label{sub:sourceStrength}

Calculation of the detectability of binary systems via
gravitational wave emission is a standard problem in the
gravitational wave community; see, for instance,
\cite{FlanaganHughes98I,FlanaganHughes98II}, and
\cite{CutlerFlanagan94}. We review the problem here for
convenience and locality of reference.

Spin interactions and higher order post-Newtonian corrections are
neglected in this paper, although they can have an appreciable
effect on parameter error estimation. For instance,
\citet{HellingsandMoore03} have shown that the inclusion of higher
harmonics of the waveform would improve the determination of the
mass parameter $\eta$, while \citet{LangandHughes06} have shown
that spinning black holes produce a modulation in the signal that
leads to a tighter bound on the sky position of the binary.

The dimensionless gravitational wave strain produced by a
circularised binary can be written as the superposition of two
independent polarization states $h_{+}$ and $h_{\times}$.  The
polarization amplitudes can be expanded in terms of harmonics as
\begin{equation}
    h_{+,\times}(\tau) = \sum_{n} h_{+,\times}^{(n)}\exp[i n
    \Phi(\tau)],
    \label{dimlessStrain}
\end{equation}
where $\tau = t - {\hat k}\cdot{\vec x}$ locates the surface of
constant phase for a gravitational wave propagating in a direction
${\hat k}$, and $\Phi(\tau)$ is the phase of the binary orbit, as
observed at the LISA detector.  When the binary is far from
coalescence, the dominant emission is the $n = 2$ quadrupole,
which is
\begin{eqnarray}
    h_{+}(\tau) &  = & \frac{2{\cal M}_{c}\left[\pi
    f\right]^{2/3}}{D_{L}}(1 + \cos^{2}\iota)\cos\left[2
    \Phi(\tau)\right] \nonumber \\
    h_{\times}(\tau) &  = & -\frac{4{\cal M}_{c}\left[\pi
    f\right]^{2/3}}{D_{L}}\cos \iota \sin\left[2
    \Phi(\tau)\right]
    \label{polarizAmplitudes}
\end{eqnarray}
where $f$ is the fundamental quadrupole frequency,
$f=2(d\Phi/dt)/(2\pi)$. We note that $h_+$ and $h_\times$ are
still functions of the observed time $\tau$.

The response of the LISA detector to the two polarizations of a
gravitational wave from a binary is given by

\begin{equation}
    y(\tau) = F_+(\theta,\phi,\iota,\psi,\tau)h_+(\tau)
           + F_\times(\theta,\phi,\iota,\psi,\tau)h_\times(\tau)
    \label{LISAresponse}
\end{equation}

\noindent where $F_+$ and $F_\times$ are the LISA form factors
that depend on the position and orientation of the source relative
to the time-dependent LISA configuration.

\subsection{Detecting Black Hole Binaries}

One measure of the ability of the LISA detector to observe a
binary signal is the signal-to-noise ratio, defined as
\begin{equation}\ ({\rm SNR})^2 = 4
\int_{0}^{\infty} \frac{ | \widetilde{h}(f) |^2}{S_{\text{LISA}}(f)} \, df \ ,
  \label{snrDefine}
\end{equation}
where $| \widetilde{h}(f) |^2 = | \widetilde{h}_{+}(f) |^2 + |
\widetilde{h}_{\times}(f) |^2$, with $\widetilde{h}_{+}(f)$ and
$\widetilde{h}_{\times}(f)$ being the Fourier transforms of the
polarization amplitudes in equation \ref{polarizAmplitudes}, and
where $S_{\text{LISA}}(f)$ is the apparent noise level of LISA's
Standard Curve Generator (\cite{SCG}, hereafter SCG), an estimate
that averages the LISA response over the entire sky and over all
polarization states and divides the LISA instrument noise,
$S_n(f)$, by this averaged response.

Previous treatment of LISA observations of binary black hole
populations \citep{Sesana07} have employed this measure of
detectability, while others \citep{Sesana04} have used a
characteristic strain $h_{c}$, following the prescription of
\cite{Thorne300}. In this measure, the raw strain $h$ of a source
is multiplied by the average number of cycles of radiation emitted
over a frequency interval $\Delta f = 1/T_{obs}$ centred at
frequency $f$. The amplitude of the characteristic strain is then
compared directly against the 1-year averaged strain sensitivity
curve from the SCG to produce an SNR. In either case, a source is
considered detectable if the resulting SNR exceeds some standard
threshold value (typically between 5 and 10).

While interesting for planning LISA data analysis pipelines, these
SNR estimates fail to address the fact that a detection is of little 
use for comparison with astrophysical theory if the parameters
of the binary are poorly determined. In particular, unless the
masses and redshifts of the detected black holes are measured, the
observations cannot be compared with the black-hole evolution
models. A more complete analysis that incorporates the
effects of uncertainty in the binary parameters is required.

Parameter error estimation for black hole binaries detected via
gravitational wave emission has been discussed by many researchers
\citep{CutlerFlanagan94, Vallisneri08, MooreHellings02,
CrowderThesis06}. This section reviews the covariance analysis for
a {\em linear least squares} process, based on the Fisher
information matrix, the method which forms the core of the error
analysis in this paper.

Let us suppose that the LISA combined data stream consists of
discrete samples of a signal given by (Eq. \ref{LISAresponse}),
with added noise:

\begin{equation}\label{signalequation}
s_\alpha=y_\alpha(\lambda_i)+n_\alpha
\end{equation}

\noindent Here, $\lambda_i$ are the parameters of the source and
$n_\alpha$ is the noise, assumed to be stationary and Gaussian.
The probability distribution of the $\alpha$th data point is
therefore
\begin{equation}\label{datapdf}
p(s_\alpha|\lambda_i)=\frac{1}{\sqrt{2\pi{\sigma}_\alpha^2}}\times
e^{-\frac{1}{2}[s_\alpha-y_\alpha(\lambda_i)]^2/{\sigma}_\alpha^2},
\end{equation}
where ${\sigma}_\alpha$ (with Greek subscript) is the standard
deviation of the noise in the $\alpha$th data point.

The likelihood function for a particular data set, with parameters
$\lambda_i$, is the product of the probabilities (Eq.
\ref{datapdf}) for each data point. It is
\begin{equation}\label{datalikelihood}
L(s_\alpha|\lambda_j)\propto \exp{\bigg[-\sum_i\frac{1}{2}\frac{[s_\alpha-y_\alpha(\lambda_j)]^2}
{\sigma_\alpha^2}\bigg]}
\end{equation}
The set of parameters, $\hat\lambda_i$, that maximizes the
likelihood function is an unbiased estimate of the the set of
actual model parameters, $\lambda_i$. To calculate the
$\hat\lambda_i$, we assume that the differences between the
estimated values and the true values, $\Delta \hat\lambda_i\equiv
\hat\lambda_i-\lambda_i$ are small enough that
$y_\alpha(\hat\lambda_i)$ can be approximated by its first-order
Taylor series expansion about $\lambda_i$:
\begin{equation}\label{taylorsignal}
y_\alpha(\hat\lambda_k) \approx y_\alpha(\lambda_k)+\nabla_i
y_\alpha\Delta\hat\lambda_i.
\end{equation}
where $\nabla_i$ represents the partial derivative with respect to
$\lambda_i$. This first-order expansion is valid when the SNR is
high enough, and the degree of correlation between the parameters
is low enough that the resulting $\Delta\lambda_i$ are small. Using Eq.
\ref{datalikelihood} we find\footnote{It is also necessary to keep
only terms that are first order in the inverse of the SNR; see
\cite{Vallisneri08}} that the likelihood is maximized by
\begin{equation}\label{MLthetas}
\Delta\lambda_i = ({\bf F}^{-1})_{ij}
\sum_\alpha\frac{1}{2}\frac{(s_\alpha-y_\alpha)\nabla_j y_\alpha}
{\sigma_\alpha^2},
\end{equation}
where the matrix ${\bf F}$ is the {\em Fisher information matrix},
with components
\begin{equation}\label{Fisherdefine}
F_{ij} = \sum_\alpha \frac{1}{2\sigma_\alpha^2}\nabla_i y_\alpha
\nabla_j y_\alpha.
\end{equation}
The expected parameter covariance matrix is
\begin{equation}\label{Fishervariance}
\langle\Delta \lambda_i \Delta \lambda_j\rangle = F_{ij}^{-1}
\end{equation}
The standard deviations in each detected parameter, $\sigma_i$
(with Latin subscript), are given by the diagonal elements of the
covariance matrix:
\begin{equation}\label{Fishersigma}
\sigma_i^2 = F_{ii}^{-1}\quad\quad\text{(no sum over {\it i})}
\end{equation}
It is important to remember that the Fisher error estimate is
accurate only when the parameter uncertainties are small compared
to the characteristic scales of the system being fit
\citep{Vallisneri06}, a condition that is not well satisfied for
all of the binaries being modelled here. In these cases, however,
the method tends to overestimate the degree of uncertainty in
systems with a sharply-defined minimum, and the resulting error
estimates tend to be conservative.

Rather than write our own Fisher error estimation codes, we have
made use of the publicly available LISA Calculator
\citep{LC,CrowderThesis06}. The LISA Calculator uses the same
instrument noise model $S_n(f)$ that is used as input to the SCG,
and an analytic signal model given by Eqs. 3 and 4. 
It takes as input a set of source parameter
values, $\lambda_j$, and outputs a set of standard deviations
(equation \ref{Fishersigma}), $\sigma_i$, for the detected
parameter values, $\hat\lambda_i$. Each detected parameter,
$\hat\lambda_i$, is assumed to have a Gaussian probability density
with mean $\lambda_i$ and standard deviation $\sigma_i$,
\begin{equation}
   p_{D}(\hat\lambda_{i}|\lambda_i)=\frac{1}{\sqrt{2\pi\sigma_{i}^{2}}}
     \exp{\Bigg[
   {-\frac{\big[\hat\lambda_{i}-\lambda_{i}\big]^{2}}{2\sigma_{i}^{2}}}\Bigg]}.
   \
   \label{LCpdf}
\end{equation}

\section{The Error Kernel}\label{sec:errorkernel}

The output of the astrophysical models of interest can be
described in terms of a coalescence rate, $\Gamma(M,\eta,z)$, per
unit redshift and time. However, the coalescence rate observed in
the LISA detector, $\Gamma'(\hat M,\hat\eta,\hat z)$, will differ
from $\Gamma(M,\eta,z)$ because some sources will be too weak to
be detected and because errors in the LISA parameter determination
will assign incorrect parameters to the source, due to the effect
of the noise on the estimation process. The effect of these errors
is summed up in the LISA error kernel, $K$:

\begin{eqnarray}\label{kerneldefinition}
\Gamma'(\hat M,\hat\eta,\hat z)&=&\int_\text {pop}
\Gamma(M,\eta,z)\times\epsilon(M,\eta,z)\nonumber \\ &
&\quad\times K(\hat M,\hat \eta,\hat z | M,\eta,z) \ dz \ d\eta \
dM , \end{eqnarray}

\noindent where $\epsilon(M,\eta,z)$ is the average detectability
of a source with parameters $\{M,\eta,z\}$ in the LISA detector.

\subsection{Calculating the Error Kernel}

As discussed in Section 3.1, coalescence rates given by the various
models are functions of the {\em population} parameters only. They
do not depend on the {\em sample} parameters, which arise from the
random relationship between the observer and a particular binary
in the population. We therefore produce a Monte Carlo average or
`marginalisation' over the sample parameters, compiling the Fisher
matrix error estimates into an `Error Kernel' which is a function
of sample parameters only.

Since we have no {\it a priori} reason to expect inhomogeneous or
anisotropic distribution, the values of the extrinsic sample
parameters of a black hole binary are assumed to be uniformly
distributed -- angular location and orientation variables are
uniformly distributed on the sky, and $\Phi_0$ is uniformly
distributed over the interval $[0,2\pi]$ (see Table
\ref{tab:parameters}).

The appropriate distribution to use for $t_{c}$ is somewhat more
complicated, owing to two primary considerations.  First,
astrophysical models of the MBH population are usually expressed
in terms of the number of coalescences per unit time and redshift.
Second, the LISA detectability of a binary (generally related to
SNR) is a monotonically decreasing function of the binary's
$t_{c}$ (all else being equal) for $t_{c}$ longer than the LISA
observation time $T_{obs}$.  Because of their relatively stronger
signals, binaries coalescing inside or soon after the LISA
observation lifetime window are far more likely to be detected
than those with long times to coalescence, so those with long
times to coalescence represent a negligible fraction of the set of
detected binaries. In calculating the error kernel, we found that the number of additional detected sources coalescing in the year following the end of LISA observation was a negligible fraction of the total, and decided to simply use a 1 year observation time and range of tc for most of our results, multiplying by 3 to get results for a 3 year observation. We found that the results were not significantly different from, for instance, using a full 3 year observation time and 4 year range of tc.

While the binary masses are also population parameters, the
models we have found in the literature generally divide their mass
spectra into very wide logarithmic bins or give no mass spectra at
all. Since the mass also has a significant effect on the detectability and parameter estimation error of a source, however, we cannot simply generate masses completely at random or treat them in some other trivial fashion. We  therefore partially marginalize over the mass and attempt to make a reasonable choice for the mass distribution within the mass bins published in the literature. This choice of mass distribution is somewhat problematic, since additional information on the mass distribution is available for some of the models studied (\cite{VHM03}, for instance), but not for others. Even if useful information on the mass distribution were available for all of the models studied, we prefer not to tailor intra-bin mass distributions to each particular model, because we want the error kernels to be model-independent. We have decided, for purposes of this paper, to use a simple uniform logarithmic
distribution within each mass bin. While this distribution can produce significant differences in detection rates for the coarsely-binned models studied here, our opinion is that it remains a reasonable choice for a model independent intra-bin mass distribution, and such problems are best solved by increasing the mass resolution of the reported model results. For similar reasons, we also completely marginalise over mass ratio. We use the three mass ranges found in \citet{Sesana07} for our mass bins. The redshift is not marginalised, and separate
Monte Carlo runs are made at uniformly spaced values of $z$ ranging
from $0\dots 20$ (see Table \ref{tab:parameters}) and the
statistics are collected as a function of redshift. The
`population' parameter space is thus considered as a collection of
two-dimensional volume elements chosen from the three mass bins
and up to 80 bins in redshift.

\begin{table}
   \centering
   \begin{tabular}{@{} ccl @{}} 
      \toprule
      \cmidrule(r){1-2} 
      Binary Parameter    & Marginalisation Range(s)\\
      \midrule
      \rowcolor{mygrey} $\psi$      & $0:\pi$&  \\
     \rowcolor{mygrey} $ \iota$       & $0:\pi/2$&  \\
      \rowcolor{mygrey}$\theta$       & $0:2\pi$& \\
      \rowcolor{mygrey}$\phi$       & $-\pi/2:\pi/2$&  \\
      \rowcolor{mygrey}$\Phi_0$  &  $0:\pi$&  \\
      \rowcolor{mygrey}$t_c$ & 0:4 years& \\
      $\eta$ & $0.0025:0.25$ \\
      $z$ & Not Marginalised\\
      \multirow{3}{*}{$ M_{tot}$} & $250 :10^4 M_\odot$ & Low Mass Case\\
      & $10^4  : 10^6 M_\odot$ & Medium Mass Case\\
      & $10^6:10^8 M_\odot$ &  High Mass Case\\

      \bottomrule
   \end{tabular}
   \caption{ This table of parameters lists the completely marginalised case parameters (grey boxes) along with their ranges and describes the treatment of the three population parameters.  Reduced mass ratio $\eta$ is averaged, $z$ is held constant and total mass ($M_{tot}$) is divided into three units. }
   \label{tab:parameters}
\end{table}

Within each volume element, values for the marginalised parameters
are chosen randomly and a covariance error analysis is performed,
with the probability distribution function (PDF) for each sample being
calculated using Eq. \ref{LCpdf}.  A typical single PDF for one sample
is shown in Figure 1a. PDFs from the random sampling of the source parameters are
stored and averaged, producing a PDF for the population parameters
corresponding to the source volume element chosen.  An example of
a marginalised PDF is shown in Figure 1b. This conditional
probability -- the probability of LISA assigning a particular
redshift to a source, given that the source parameters are within
this mass and redshift bin -- is exactly what is meant by the
error kernel in Eq. \ref{kerneldefinition}. The marginalised error
kernel is thus

\begin{equation}
   K(\hat\lambda_{i}|\lambda_{i})=\frac{1}{N}\sum_{\mu=1}^{N}\frac{1}{\sqrt{2\pi\sigma_{i,\mu}^{2}}}
   \exp{\Bigg[{-\frac{\big[\hat\lambda_{i}-\lambda_{i}\big]^{2}}{2\sigma_{i,\mu}^{2}}}\Bigg]},
   \label{LISApsf}
\end{equation}
where $\sigma_{i,\mu}$ is the uncertainty in the $i$th parameter
in the $\mu$th randomly generated set of source parameters within
the bin. $N$ is the total number of samples generated in the bin.

\begin{figure*}
   \centering
   \includegraphics[width=1\textwidth]{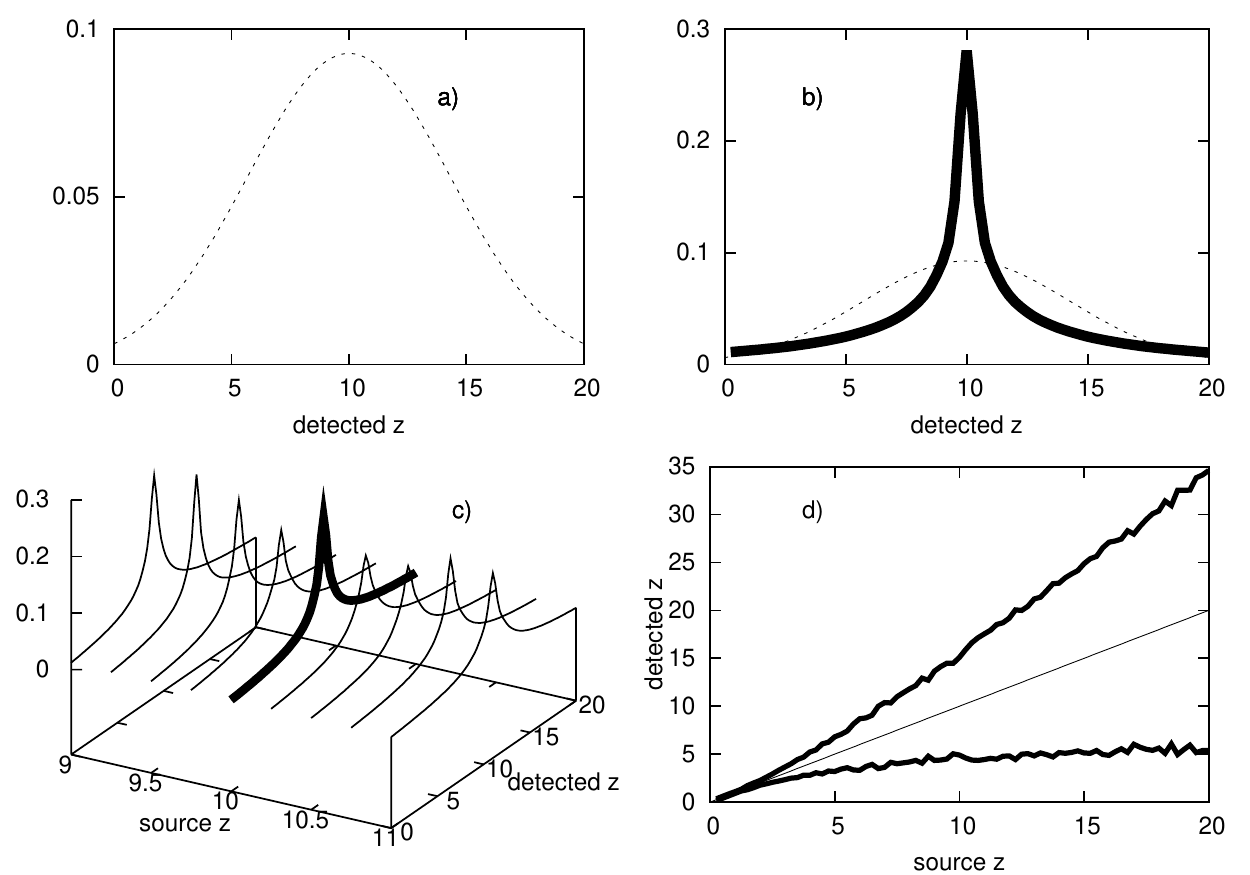}
   \caption{Creating the error kernel for the `medium mass' bin and marginalised parameters described in Table \ref{tab:parameters}, total masses between $10,000 M_\odot$ and $100,000 M_\odot$.  a) The Gaussian PDF implied by simple RMS error.   b) Adding the probability densities resulting from many different marginalised parameters results in a highly non-normal distribution.  Shown here (in solid black) is the distribution of possible detections given several hundred sources at a redshift of 10 in a mass range $10^4 : 10^6 M_\odot$.  Overlaid (in dashes) is the distribution obtained by simply adding the errors in quadrature.  c) The $z_s=10$ PDF inserted into its place in the error kernel.  Source redshifts are sampled at even redshift intervals of 0.25. d) 70\% confidence intervals for LISA determination of redshift gives an overview of the resulting error kernel.  }
   \label{fig:kernel_process_figure}
\end{figure*}

Although each individual element of the sum is a Gaussian, the
process results in a non-Gaussian distribution since the size of
the uncertainty, $\sigma_{i,\mu}$, changes with each new set of sample
parameters. This can be seen by comparing Figures
\ref{fig:kernel_process_figure}a and
\ref{fig:kernel_process_figure}b. The sum of Gaussians, all
centred on the value of the source parameter, has a taller peak
and fatter tails than a single Gaussian with average standard
deviation $\big[\sum_{\mu=1}^{N}\sigma_{i,\mu}^2\big]^{0.5}/N$.

There are two limitations in the way we have generated the error
kernel that stem from our use of the linear least-squares LISA
Calculator. First, sources at moderate redshift and large sigma
will have tails that extend to low $z$, even though a true nearby
source would not be confused with a stronger source at moderate
redshift. Production of the true PDF for such a case would involve a more
complete algorithm, avoiding the limitations of linear
least-squares analysis and resulting in a shorter low-z tail.
Second, the LISA Calculator, like many least-squares
tools, drops an ill-determined parameter when the information
matrix is singular, and gives an inappropriately low sigma for the
remaining parameters. This did occur in a number of the cases
we ran and contributed some anomalously strong peaks to the Monte
Carlo averaging.

\begin{figure} 
    \centering
    \includegraphics[width=\columnwidth]{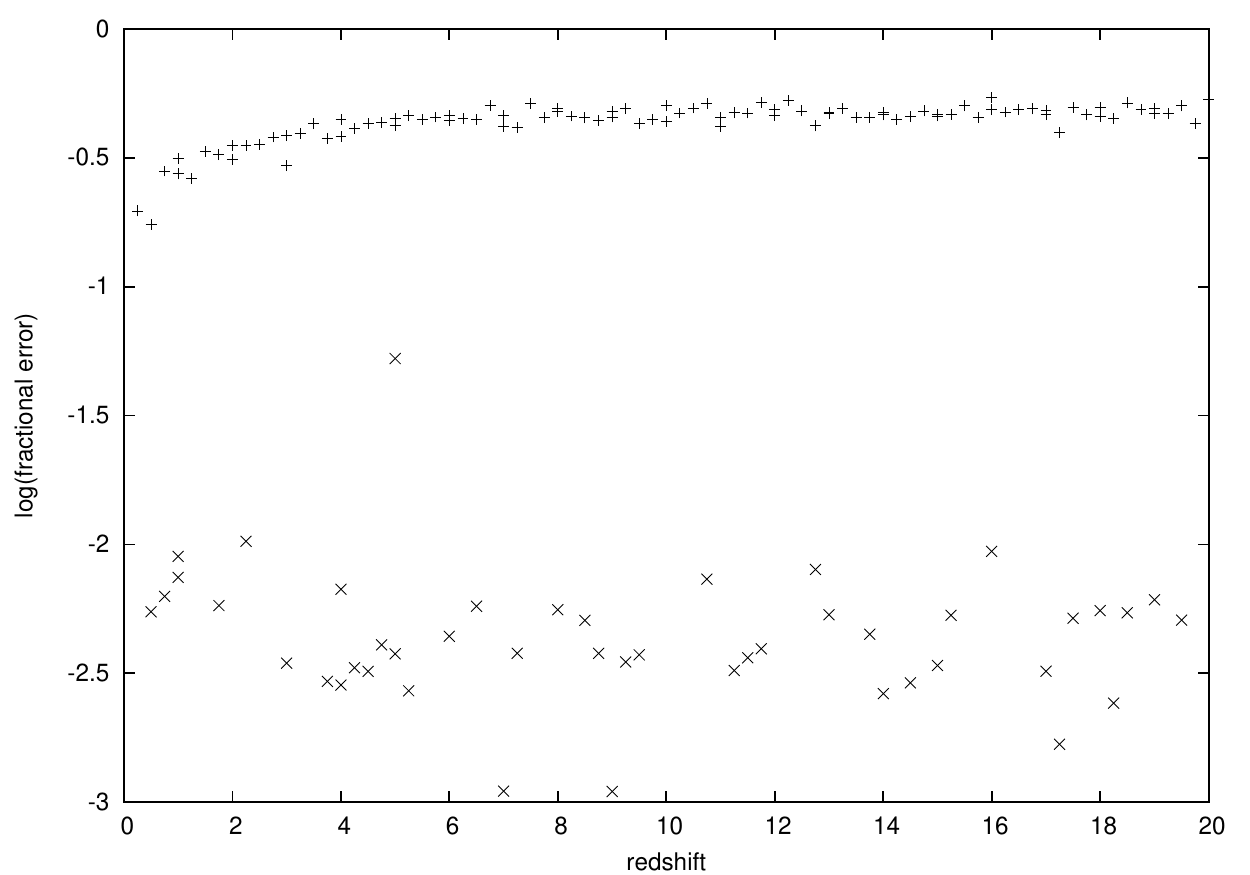}
    \caption{Average fractional error for binaries with total mass between $10,000 M_\odot$ and $1,000,000 M_\odot$ and reduced mass between $0.001$ and $0.25$, using $+$ for redshift and x for reduced mass. Chirp mass errors are not shown, but are an order of magnitude below the reduced mass errors. Error in redshift dominates the reduced mass error by over an order of magnitude.}
    \label{fig:errors}
 \end{figure}

Our covariance studies found that the fractional
uncertainties in the redshifted mass variables were always much less than the
fractional uncertainties in redshift (see figure
\ref{fig:errors}), and are insignificant compared to our coarse
mass binning\footnote{Rest-frame variables with dimensions of mass will have an error that is 100\% correlated with the error in the redshift, a detail we will investigate in future work.}. 
For the purposes of this initial study, we have ignored the mass errors and considered
only the distribution of the detected redshifts. The error kernel is thus
 \begin{equation}
 K_i(\hat z,z) = \frac{1}{N_i}\sum_{\mu_i}\frac{1}{\sqrt{2\pi\sigma_{z,\mu_i}^2}}\exp\left[-\frac{(\hat z-z)^2}{2\sigma_{z,\mu_i}^2}\right],
 \label{eq:LISAKernel1}
 \end{equation}
where $i$ corresponds to one of the mass ranges defined in Table
\ref{tab:parameters} and where the sigmas are understood to be the
uncertainties determined for each mass bin. Thus, our Monte Carlo
study returns three error kernels at a given redshift, one for 
each mass range. Each error kernel consists of a set of PDFs, one 
PDF for each redshift bin $\hat z$. These source redshifts are uniformly 
spaced between 0 and 20 (see figure \ref{fig:kernel_process_figure}c). 
Each of these PDFs represent Monte Carlo averages of LISA calculator
Gaussian PDFs, varied over the {\it sample} parameters and the
mass parameter ranges listed in Table \ref{tab:parameters}. Each
error kernel contains about 2 million LISA calculator runs. One
may get a feel for the shape of the entire kernel by looking at
the $70\%$ confidence intervals shown in Figure
\ref{fig:kernel_process_figure}d.

\subsection{Applying the Error Kernel}

Once the error kernel in Eq. \ref{eq:LISAKernel1} has been
calculated, it may be applied to any population model that gives
the source coalescence rate in the corresponding mass ($i$) and
redshift bin $\Gamma_i(z)$, producing a prediction for the
detected coalescence rate, $\Gamma_i'(\hat z)$. This is
accomplished by a straightforward convolution, which we write in
continuum form as
\begin{equation}\label{kernelconvolution}
\Gamma_i'(\hat z)=\int_0^{z_{\rm max}} \Gamma_i(z)\times K_i(\hat z,z) dz.
\end{equation}

As we noted in Eq. \ref{kerneldefinition}, some of the binaries
sampled will have signal-to-noise ratios too small to be
detectable, regardless of the error in the parameters. Previous
analysis by \citet{Sesana07} has used the fiducial SNR limit of 5.
We chose to use a cutoff at ${\rm SNR}=8$, but the number of
additional sources dropped due to our more conservative cutoff was
negligible. The PDFs of binaries which do not make the SNR cutoff
should not be included in the error kernels, but the proportion of
rejected binaries in each source parameter bin must still be taken
into account when the error kernels are applied to the models.
Therefore, the error kernels (which would otherwise normalise to
1) are weighted with the fraction of binaries in each bin,
$\epsilon_i(z)$, having ${\rm SNR} \geq 8$. Taking this
detectability into account, the convolved coalescence rates may be
written
\begin{equation}\label{eq:kernelconvolution_with_epsilon}
\Gamma_i'(\hat z)=\int_0^{z_{\rm max}} \Gamma_i(z)\times\epsilon_i(z)\times K_i(\hat z,z) dz.
\end{equation}

\section{Discriminating Between Population Models}\label{sec:modelcomparison}

As an illustration of LISA's ability to discriminate between black
hole population models, we consider four formation
models. The models we have chosen for the demonstration -- those
by \citet{VHM03}, \citet{KBD}, and two by \citet{BVR}, one
with `high' feedback and one with `low' feedback, (hereafter
VHM, KBD, BVRhf, and BVRlf respectively) -- are variations on the
extended Press-Schecter (EPS) formalism by \citet{EPS} which
assigns a mass-dependent probability to halo mergers. Key variations between 
these models are their assumptions for accretion, their binary
hardening scenarios, their choices for mass and redshift of seed
formation, and the details of the way they handle MBH binary
interactions near the merger.

\subsection{Convolving The Models with the LISA Error Kernel}

\begin{figure} 
   \centering
   \includegraphics[width=1\columnwidth]{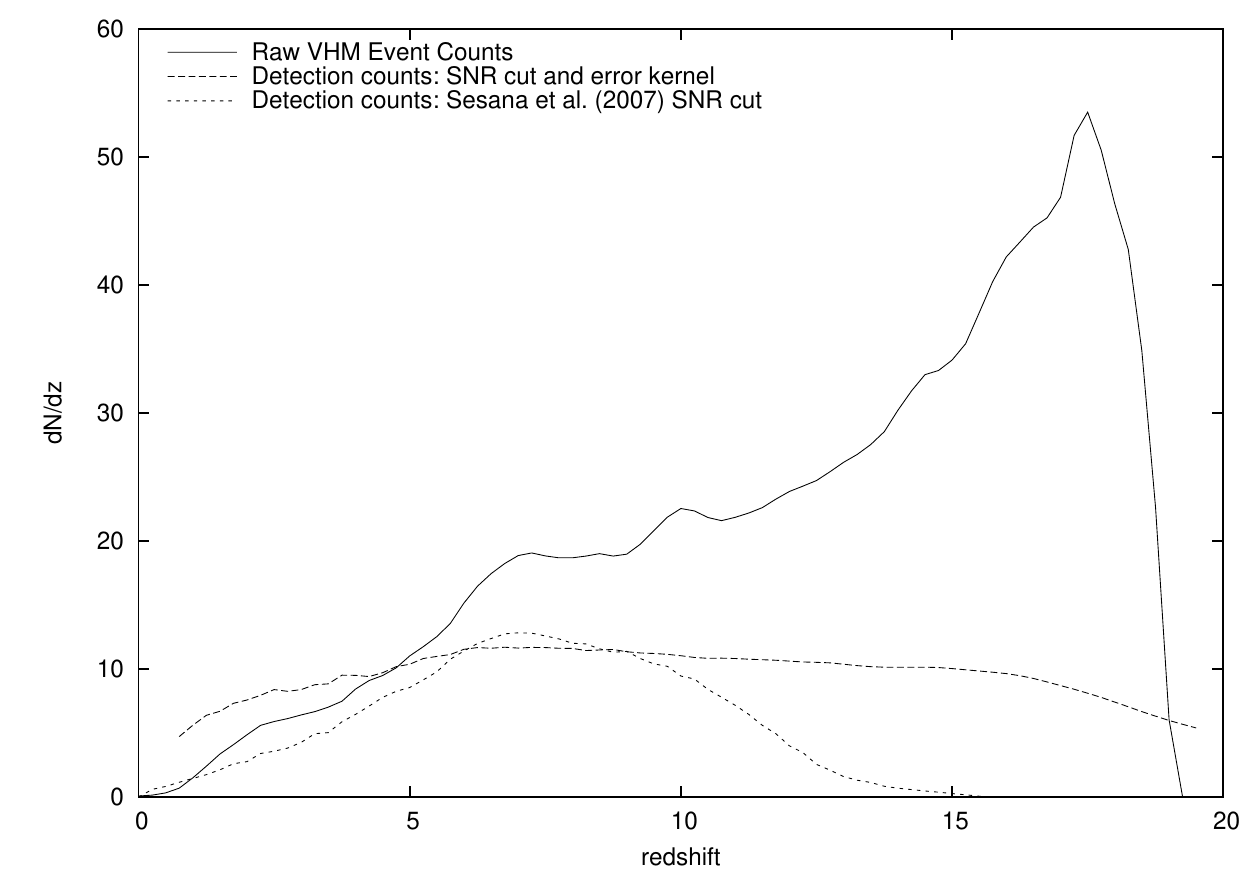}
   \caption{The effect of the LISA error kernel on the MBH binary population predicted by \citet{VHM03} for a 3 year observation.  Solid line: modelled source distribution for masses below $10^8 M_\odot$. Dotted line: modelled source distribution with $\mathrm{SNR}>8$ cut, as applied by \citet{Sesana07}. Dashed line: convolution of modelled source distribution with LISA error kernel. The error kernel distribution incorporates the large errors inherent to binary redshift determination.}
   \label{fig:VHM_total_mass_compare}
\end{figure}

The effect of the error kernel on population model testing can be
seen in Figure \ref{fig:VHM_total_mass_compare}, where we have
taken the VHM model with its three mass bins lumped together,
spanning the range from 300 $M_\odot$ to $10^8M_\odot$. The VHM
model, with its unique seeding scenario,  predicts a large number
of low mass, high redshift binaries. The distribution, shown as
the large-amplitude solid curve in Figure
\ref{fig:VHM_total_mass_compare}, peaks and then rolls off sharply
at $z\approx 17$. The number of sources in this model that are
expected to be visible with LISA, using a SNR$>8$ cut-off relative
to the averaged sensitivity curve from the SCG, is given by the
low-amplitude dotted curve \citep[see][]{Sesana07}. The dashed curve represents our results,
produced by integrating the model with the kernel, as in Eq.
\ref{kernelconvolution}. It gives predictions for the distribution of best-fitting parameters detected by LISA, using a cutoff at SNR $=8$. The obvious point to be made is that the
redshift uncertainties smear out a model's features, so that,
while the VHM model of Figure \ref{fig:VHM_total_mass_compare} has a very distinctive shape in
its theoretical incarnation, the distribution that would be observed by LISA will be far less so.

The presence of detected sources at very low ($z<2$) redshifts, where the VHM population model says that there should be very few, is a result of the excessively large low-z tails of the error kernel discussed in Section \ref{sec:errorkernel}. The less sharply peaked shape of the error kernel results as compared with the Sesana results, however, is a consquence of applying detection errors to the redshifts, and will persist even when more robust error-estimation techniques are employed.

Even with our SNR cutoff, we predict more visible sources (179 vs. 96) than did \citet{Sesana07}. This is due in part to our error kernel, with its use of the logarithmic mass distributions within the large mass bins found in the literature, having more massive binaries at high redshift than is actually the case for the VHM models. We also found a puzzling discrepancy in \cite{Sesana07} between their stated event counts (~250 for the VHM model with a 3 year range of coalescence times) and the event counts found by integrating the curves in their figure 1 (over 400 for the same case). Since our model event rates were obtained by extracting the curves from that figure, this discrepancy could also contribute to the differing number of visible sources.

\subsection{Discriminating Between Models}

\begin{figure*} 
   \centering
  \resizebox{\textwidth}{!}{\input{figures/LISA_Quadplot2.tex}}
   \caption{Binary merger redshift spectra, after smearing by the LISA error kernel in the style of \citet{Sesana07}. Solid line - VHM, short-long dashed line - KBD, short dashed line - BVRlf, dashed line - BVRhf.}
    \label{fig:LISA_Quadplot}
\end{figure*}
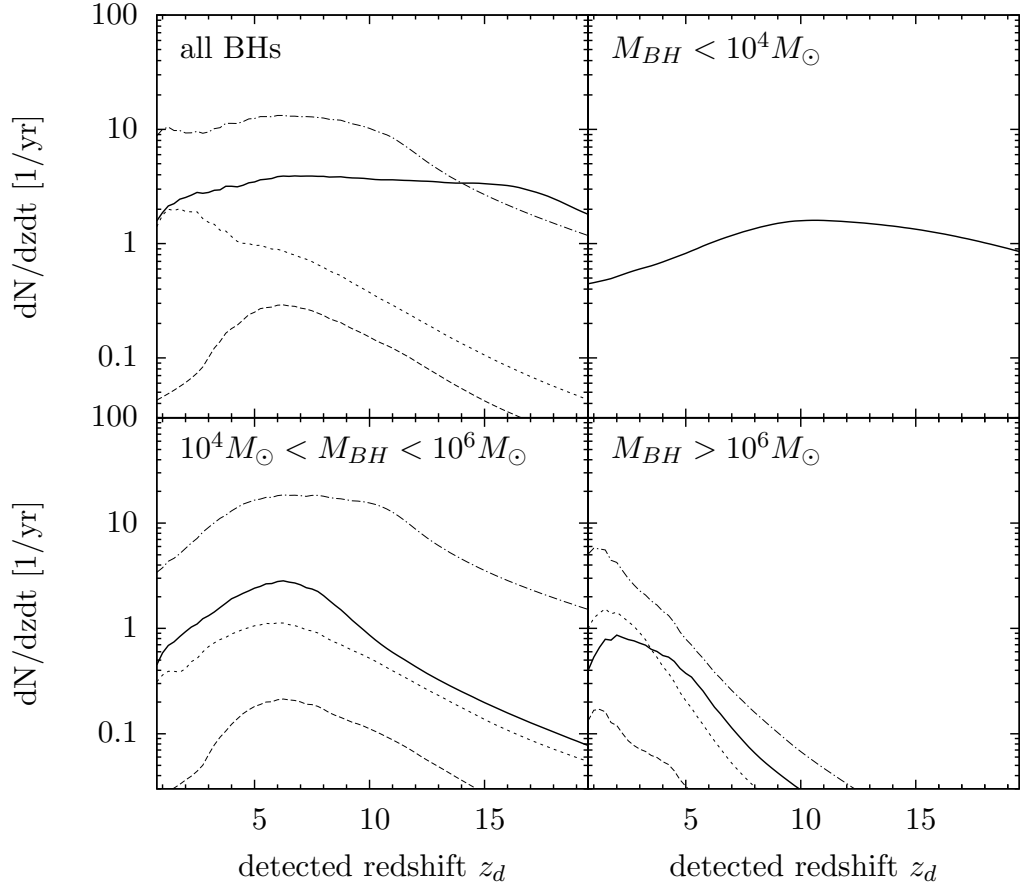

For the four models we have chosen to consider as
illustrative examples, the results of the error kernel
convolutions are shown in Figure \ref{fig:LISA_Quadplot}. The
graph in the upper left is for all masses and the other three
graphs represent the three mass bins we used. We use a
modified version of the Kolmogorov-Smirnov (K-S) test as a measure of separability of the
models. Our test differs from the K-S test in that it is sensitive to differences in the model event rates as well as to the cumulative distribution functions (CDFs) of samples drawn from the models. For each pair of models shown in Table \ref{tab:model
comparison}, we have simulated Monte Carlo draws of the number of
sources in each redshift bin, using a Poisson distribution with
probability given by each of the models. One thousand draws were
taken for each model and our test statistic was calculated,
finding the greatest deviation between the cumulative
histograms of the two models. The probability that the two
draws were from the same model was then found by Monte Carlo sampling from the null hypothesis that the two samples have the same CDF and have event counts which are Poisson distributed with identical rate parameters.

Several comparisons were done between the four models chosen from \cite{Sesana07}.
For each comparison, we assumed one year of LISA observations and
randomly drew coalescence parameters using the probability
distributions for the two models being compared. In each set of
draws, our modified K-S statistic, $E$, was determined and $Q$, the probability
that two such draws would be produced by the same model, was calculated. The
results for each random draw were then averaged over one thousand
such realisations, giving the values displayed in Table
\ref{tab:model comparison}. In the two data columns, we show the
results from the raw models themselves, with no parameter
uncertainties taken into account, and the results from the models
after they have been convolved with the LISA error kernel. As can
be seen in the table, the probability that any of the simulated
data sets for one model might have been produced by one of the
other models is small. The models examined here appear to be
easily distinguishable from each other, with the exception of the comparison of the BVRhf model with the BVRlf model with its average $Q$ value of 0.055 (corresponding to a rather shaky $94.5\%$ confidence). Even in that case, when we look at the median of $Q$ rather than its mean, we find it to be $0.012$ ($98.8\%$ confidence), implying that a BVRlf realization can usually be distinguished from a BVRhf realization.

\begin{table}
   \centering
   \begin{tabular}{@{} l|cc|cc>{\bfseries}cc @{}} 

      Models Compared & \multicolumn{2}{c|}{Before LISA kernel} & \multicolumn{2}{c}{After LISA Kernel} \\
\hline
            &$\langle E\rangle$&$\langle Q\rangle$ &$\langle E\rangle$&$\langle Q\rangle$\\
      VHM - KBD   &  175.6&$<10^{-4}$ & 90.5&$<10^{-4}$ \\
      VHM - BVRlf       & 119.7 &$<10^{-4}$  & $49.7$&$<10^{-4}$ \\
      VHM - BVRhf      &  132.9&$<10^{-4}$ & 58.9&$<1.0\times10^{-4}$ \\
      BVRhf - BVRlf  & 14.29 & $0.021$ & 9.82  &  0.055 \\
   \end{tabular}
   \caption{Comparisons of models before and after convolution with LISA kernel, for binaries coalescing within the observation window, assuming one year of observation time. The `Before LISA' comparisons effectively assume that all of the sources are detectable and have zero redshift error, while the `After LISA' comparisons incorporates the effects of both parameter uncertainty and detectability. $E$ is the maximum deviation between the cumulative histograms of random draws from the two models. $Q$ is the corresponding probability that random fluctuations could be responsible for the deviation.}
   \label{tab:model comparison}
\end{table}

\section{Conclusion}\label{sec:conclusion}

In this paper, we have constructed an `error kernel' for the
LISA detector, using a public-domain error computation module in a
Monte-Carlo data pipeline, marginalising over the `sample'
parameters. This result represents a first implementation of a
model-level error function for a gravitational wave observatory, a
concept which is central to all types of astronomy. The error
kernel approach introduced here is designed to replace simple SNR cuts
as the interface between population modellers and gravitational
wave signal specialists.

The error kernel can be computed as a function of any of the
population parameters. We have chosen to use redshift alone, but
total mass or mass ratio could easily have been added. Once
calculated, contours of the error kernel can be used to visualise
the relative impact of the detector noise, analysis technique, and
choice of parameter binning on the ability of the detector to
determine the parameters of the sources in the model.

We reiterate that we are limited by working with the models
as published, with their emphasis on redshift as the parameter of
interest, rather than having access to more detailed model results giving
populations as functions of masses and redshifts. In particular, while the logarithmic intra-bin mass distribution is as reasonable a choice as any other, the mass bins published in the literature are so large that no model-independent distribution can produce an error kernel that is free from significant bias compared to what would be obtained using the full model results. Furthermore, with LISA's exquisite resolution in the redshifted mass variables, the mass distribution is likely to provide useful astrophysical information in its own right. It is our opinion that a mass resolution of at least 1 bin per decade over the range $10^2\dots 10^6 M_{\odot}$ is necessary to meaningfully specify the mass dependence of these populations.
We are currently working toward improved comparisons using more detailed models.

The demonstration in this paper for four model
comparisons looks at the distinguishability of the models based on
what might eventually be a catalogue of LISA binaries. Our approach
has been to do a forward modelling from the relevant parameters of
the population model to the detected parameters of sources seen in
the LISA data. The fact that those predictions are statistically
different suggests that LISA data will have appreciable model-discriminating power. 

When the LISA data are finally available, the analysis will include
backward, Bayesian, modelling which calculates and interprets the contribution 
of each detected source to a likelihood function for the models being tested. 
This backward population analysis framework will be an essential component
of future LISA data analysis, and is a natural direction of our future 
work in the long term.

\vspace{+0.5cm} The work presented here was supported in part by
NASA EPSCoR \#437259.  The work of SLL was supported in part by
NASA award NNG05GF71G. We extend our thanks to our anonymous referee, whose comments have been of much help in clarifying portions of this paper.

\bibliographystyle{mn}

\bibliography{mbh21}

\end{document}

%% file: figures/LISA_Quadplot2.tex
\begingroup
  \makeatletter
  \providecommand\color[2][]{%
    \GenericError{(gnuplot) \space\space\space\@spaces}{%
      Package color not loaded in conjunction with
      terminal option `colourtext'%
    }{See the gnuplot documentation for explanation.%
    }{Either use 'blacktext' in gnuplot or load the package
      color.sty in LaTeX.}%
    \renewcommand\color[2][]{}%
  }%
  \providecommand\includegraphics[2][]{%
    \GenericError{(gnuplot) \space\space\space\@spaces}{%
      Package graphicx or graphics not loaded%
    }{See the gnuplot documentation for explanation.%
    }{The gnuplot epslatex terminal needs graphicx.sty or graphics.sty.}%
    \renewcommand\includegraphics[2][]{}%
  }%
  \providecommand\rotatebox[2]{#2}%
  \@ifundefined{ifGPcolor}{%
    \newif\ifGPcolor
    \GPcolorfalse
  }{}%
  \@ifundefined{ifGPblacktext}{%
    \newif\ifGPblacktext
    \GPblacktexttrue
  }{}%
  \let\gplgaddtomacro\g@addto@macro
  \gdef\gplbacktext{}%
  \gdef\gplfronttext{}%
  \makeatother
  \ifGPblacktext
    \def\colorrgb#1{}%
    \def\colorgray#1{}%
  \else
    \ifGPcolor
      \def\colorrgb#1{\color[rgb]{#1}}%
      \def\colorgray#1{\color[gray]{#1}}%
      \expandafter\def\csname LTw\endcsname{\color{white}}%
      \expandafter\def\csname LTb\endcsname{\color{black}}%
      \expandafter\def\csname LTa\endcsname{\color{black}}%
      \expandafter\def\csname LT0\endcsname{\color[rgb]{1,0,0}}%
      \expandafter\def\csname LT1\endcsname{\color[rgb]{0,1,0}}%
      \expandafter\def\csname LT2\endcsname{\color[rgb]{0,0,1}}%
      \expandafter\def\csname LT3\endcsname{\color[rgb]{1,0,1}}%
      \expandafter\def\csname LT4\endcsname{\color[rgb]{0,1,1}}%
      \expandafter\def\csname LT5\endcsname{\color[rgb]{1,1,0}}%
      \expandafter\def\csname LT6\endcsname{\color[rgb]{0,0,0}}%
      \expandafter\def\csname LT7\endcsname{\color[rgb]{1,0.3,0}}%
      \expandafter\def\csname LT8\endcsname{\color[rgb]{0.5,0.5,0.5}}%
    \else
      \def\colorrgb#1{\color{black}}%
      \def\colorgray#1{\color[gray]{#1}}%
      \expandafter\def\csname LTw\endcsname{\color{white}}%
      \expandafter\def\csname LTb\endcsname{\color{black}}%
      \expandafter\def\csname LTa\endcsname{\color{black}}%
      \expandafter\def\csname LT0\endcsname{\color{black}}%
      \expandafter\def\csname LT1\endcsname{\color{black}}%
      \expandafter\def\csname LT2\endcsname{\color{black}}%
      \expandafter\def\csname LT3\endcsname{\color{black}}%
      \expandafter\def\csname LT4\endcsname{\color{black}}%
      \expandafter\def\csname LT5\endcsname{\color{black}}%
      \expandafter\def\csname LT6\endcsname{\color{black}}%
      \expandafter\def\csname LT7\endcsname{\color{black}}%
      \expandafter\def\csname LT8\endcsname{\color{black}}%
    \fi
  \fi
  \setlength{\unitlength}{0.0500bp}%
  \begin{picture}(8640.00,6048.00)%
    \gplgaddtomacro\gplbacktext{%
      \csname LTb\endcsname%
      \put(1188,823){\makebox(0,0)[r]{\strut{} 0.1}}%
      \put(1188,1557){\makebox(0,0)[r]{\strut{} 1}}%
      \put(1188,2290){\makebox(0,0)[r]{\strut{} 10}}%
      \put(1188,3023){\makebox(0,0)[r]{\strut{} 100}}%
      \put(2000,220){\makebox(0,0){\strut{} 5}}%
      \put(2800,220){\makebox(0,0){\strut{} 10}}%
      \put(3600,220){\makebox(0,0){\strut{} 15}}%
      \put(418,1731){\rotatebox{90}{\makebox(0,0){\strut{}dN/dzdt [1/yr]}}}%
      \put(2820,-110){\makebox(0,0){\strut{}detected redshift $z_d$}}%
      \put(1480,2802){\makebox(0,0)[l]{\strut{}$10^4 M_\odot<M_{BH}<10^6 M_\odot$}}%
    }%
    \gplgaddtomacro\gplfronttext{%
    }%
    \gplgaddtomacro\gplbacktext{%
      \csname LTb\endcsname%
      \put(4188,823){\makebox(0,0)[r]{\strut{}}}%
      \put(4188,1557){\makebox(0,0)[r]{\strut{}}}%
      \put(4188,2290){\makebox(0,0)[r]{\strut{}}}%
      \put(4188,3023){\makebox(0,0)[r]{\strut{}}}%
      \put(5000,220){\makebox(0,0){\strut{} 5}}%
      \put(5800,220){\makebox(0,0){\strut{} 10}}%
      \put(6600,220){\makebox(0,0){\strut{} 15}}%
      \put(5820,-110){\makebox(0,0){\strut{}detected redshift $z_d$}}%
      \put(4480,2802){\makebox(0,0)[l]{\strut{}$M_{BH}>10^6 M_\odot$}}%
    }%
    \gplgaddtomacro\gplfronttext{%
    }%
    \gplgaddtomacro\gplbacktext{%
      \csname LTb\endcsname%
      \put(4188,3440){\makebox(0,0)[r]{\strut{}}}%
      \put(4188,4236){\makebox(0,0)[r]{\strut{}}}%
      \put(4188,5031){\makebox(0,0)[r]{\strut{}}}%
      \put(4188,5827){\makebox(0,0)[r]{\strut{}}}%
      \put(5000,2804){\makebox(0,0){\strut{}}}%
      \put(5800,2804){\makebox(0,0){\strut{}}}%
      \put(6600,2804){\makebox(0,0){\strut{}}}%
      \put(4480,5587){\makebox(0,0)[l]{\strut{}$M_{BH}<10^4 M_\odot$}}%
    }%
    \gplgaddtomacro\gplfronttext{%
    }%
    \gplgaddtomacro\gplbacktext{%
      \put(1188,3440){\makebox(0,0)[r]{\strut{} 0.1}}%
      \put(1188,4236){\makebox(0,0)[r]{\strut{} 1}}%
      \put(1188,5031){\makebox(0,0)[r]{\strut{} 10}}%
      \put(1188,5827){\makebox(0,0)[r]{\strut{} 100}}%
      \put(2000,2804){\makebox(0,0){\strut{}}}%
      \put(2800,2804){\makebox(0,0){\strut{}}}%
      \put(3600,2804){\makebox(0,0){\strut{}}}%
      \put(418,4425){\rotatebox{90}{\makebox(0,0){\strut{}dN/dzdt [1/yr]}}}%
      \put(1480,5587){\makebox(0,0)[l]{\strut{}all BHs}}%
    }%
    \gplgaddtomacro\gplfronttext{%
    }%
    \gplbacktext
    \put(0,0){\includegraphics{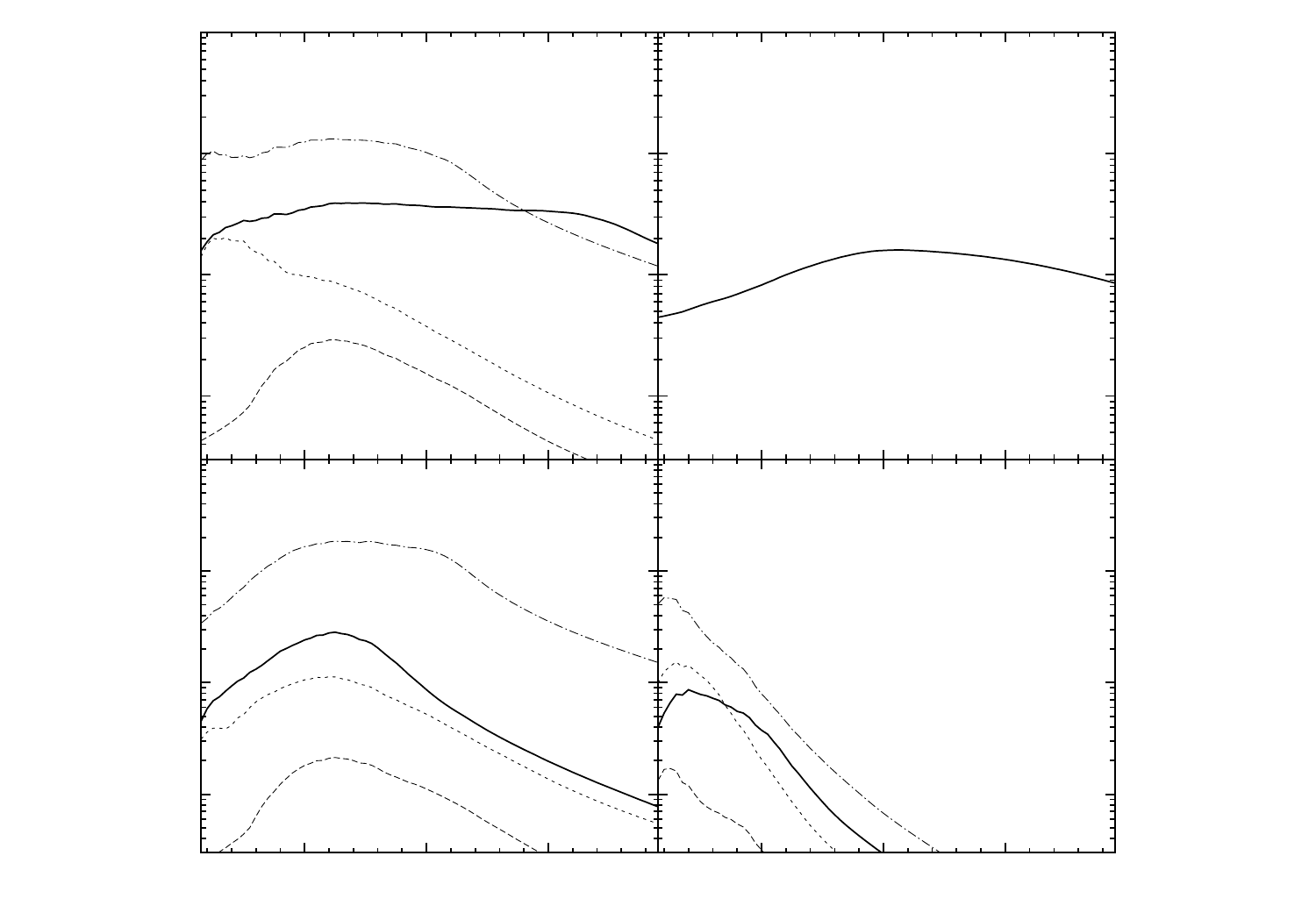}}%
    \gplfronttext
  \end{picture}%
\endgroup